%% file: template.tex
\documentclass{article}

\usepackage{arxiv}

\usepackage[utf8]{inputenc} 
\usepackage[T1]{fontenc}    
\usepackage[hidelinks]{hyperref}       
\usepackage{url}            
\usepackage{booktabs}       
\usepackage{amsmath}
\usepackage{amsfonts}       
\usepackage{nicefrac}       
\usepackage{microtype}      
\usepackage{lipsum}		
\usepackage{graphicx}
\usepackage{cite}
\usepackage{doi}

\title{Managing Context and Communication in Distributed Agentic UAV Swarms\thanks{This paper has been accepted for presentation at the 24th IEEE Consumer Communications \& Networking Conference 2027 (CCNC 2027)}}

\date{}

\newcommand{\authorname}[2]{%
  \makebox[0.45\textwidth][c]{\href{https://orcid.org/#1}{\includegraphics[scale=0.06]{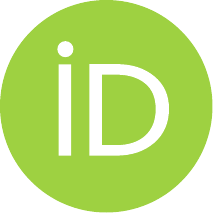}\hspace{1mm}#2}}%
}

\author{\authorname{0009-0009-1040-2138}{Andrea Iannoli}\\
	Department of Computer Science and Engineering\\
	University of Bologna\\
	Bologna, Italy \\
	\texttt{andrea.iannoli@unibo.it} \\
	\And
	\authorname{0000-0003-4936-9645}{Ivan Zyrianoff}\\
	Department of Computer Science and Engineering\\
	University of Bologna\\
	Bologna, Italy \\
	\texttt{ivan.zyrianoff@unibo.it} \\
	\AND
	\authorname{0000-0002-0552-2444}{Angelo Trotta}\\
	Department of Computer Science and Engineering\\
	University of Bologna\\
	Bologna, Italy \\
	\texttt{a.trotta@unibo.it} \\
	\And
	\authorname{0000-0001-9714-3777}{Lorenzo Gigli}\\
	Technology Innovation Institute (TII)\\
	United Arab Emirates \\
	\texttt{lorenzo.gigli@tii.ae} \\
	\AND
	\authorname{0000-0001-7496-7597}{Marco Di Felice}\\
	Department of Computer Science and Engineering\\
	University of Bologna\\
	Bologna, Italy \\
	\texttt{marco.difelice3@unibo.it}
}

\renewcommand{\shorttitle}{Managing Context and Communication in Distributed Agentic UAV Swarms}

\hypersetup{
pdftitle={Managing Context and Communication in Distributed Agentic UAV Swarms},
pdfauthor={Andrea Iannoli, Ivan Zyrianoff, Angelo Trotta, Lorenzo Gigli, Marco Di Felice},
pdfkeywords={Large Language Models,
Model Context Protocol,
Web of Things,
UAV Swarm Control,
Agent-Based Reasoning},
}

\begin{document}
\maketitle

\begin{abstract}
Unmanned aerial vehicle (UAV) swarms increasingly rely on language-model agents to provide adaptive mission-level reasoning in uncertain environments. Fully distributed control, in which each UAV hosts an independent Small Language Model (SLM), removes reliance on a centralized coordinator but introduces an information-management problem: long-running interaction histories can degrade the reasoning context, while indiscriminate information dissemination increases communication and inference overhead. We address these challenges with a distributed UAV-agent architecture that enables continuous local SLM control through an event-driven reason--act--observe lifecycle. Runtime knowledge is represented as structured atomic notes and organized into core, local, and peer-specific memory. A deterministic interest-aware gossip engine selectively disseminates these notes according to recipient-specific semantic novelty and recency. We evaluate the architecture using ten UAVs in a simulated search-and-rescue mission. Our approach completes all experimental runs, whereas unrestricted flooding messages completes only 70--85\%, and delegating forwarding decisions to the SLM prevents mission completion in every run. Compared with unrestricted flooding, our approach approximately halves inference-token consumption, reduces transmitted data, and achieves lower survivor-count error.
\end{abstract}


\input{intro}
\input{related}
\input{arch}
\input{gossip}
\input{eval}
\input{conclusions}

\bibliographystyle{IEEEtran}
\bibliography{references}  






\end{document}

%% file: intro.tex
\section{Introduction}
Unmanned aerial vehicle (UAV) swarms combine rapid deployment, parallel sensing, and wide-area coverage, making them attractive for search-and-rescue, disaster recovery, environmental monitoring, and infrastructure inspection. These missions are executed in uncertain environments in which predefined plans may become inadequate as hazards, observations, connectivity, and resource availability change~\cite{sapkota2025uavs}. Agentic AI can complement conventional perception and flight-control components with goal interpretation, contextual reasoning, and adaptive planning, enabling UAVs to revise mission-level decisions when unexpected conditions arise~\cite{nguyen2026agentic}.

Embedding such reasoning capabilities into continuous UAV swarm control, however, is challenging. Model decisions must be grounded in heterogeneous sensing and actuation interfaces, executed through a long-running closed loop, and constrained by safety requirements despite the non-determinism of language models~\cite{iannoli2026say}. To mitigate these difficulties, language-level reasoning in many existing systems remains centralized: a cloud, edge, or swarm-level agent maintains a global view and generates mission plans, while individual UAVs execute them through conventional controllers~\cite{Ahmed2026LLM-Guided,iannoli2026say}. Such architectures depend on reliable connectivity to an external coordinator and introduce communication and computational bottlenecks as observations and control decisions are continuously exchanged and the swarm state grows~\cite{nguyen2026agentic}. Assigning an independent agent to each UAV instead enables local decision making, distributes the reasoning workload, and removes a central point of failure. Although distributed LLM-based UAV designs are beginning to emerge~\cite{Wang2025RALLY:}, fully decentralized control introduces additional challenges: each agent reasons from partial observations, coordinates over bandwidth-constrained links, and operates under onboard memory, computation, energy, and latency constraints that favor smaller Small Language Models (SLMs) with more limited reasoning capabilities~\cite{nguyen2026agentic}. However, existing distributed agentic UAV architectures do not explicitly address how long-running local knowledge should be organized and selectively disseminated among resource-constrained agents.

Indeed, every action produces observations and execution feedback, while cooperation contributes a continuous stream of peer messages. As these records accumulate, replaying the complete transcript can bury mission-critical information among transient or redundant events. This degradation is commonly described as \textit{context rot}, which refers to the model failing to use relevant information embedded in long inputs even before the context limit is reached~\cite{liu-etal-2024-lost}. In a UAV swarm, such degradation can directly affect mission safety and completion, e.g., an agent that overlooks a mandatory landing instruction may continue unsafe execution, potentially resulting in a collision, vehicle loss, or mission failure. Long-running control therefore requires a bounded and structured representation of accumulated knowledge that preserves mission-relevant information while separating it from transient history~\cite{packer2023memgpt,xu2025amem}. Regarding information dissemination, indiscriminately flooding observations wastes bandwidth and fills peer contexts with redundant information. Delegating each per-message, per-neighbor forwarding decision to the SLM can also introduce substantial reasoning overhead, enlarge the context, and interfere with the UAV control lifecycle.

We therefore propose a fully distributed architecture in which every UAV hosts an independent SLM-based agent, a Context Manager, and an Interest-Aware Gossip Engine. Each UAV continuously reasons over mission information, locally accumulated knowledge, execution feedback, and asynchronously received peer information, enabling its subsequent actions to adapt without a centralized coordinator. Knowledge produced during this process is represented as \textit{atomic notes}, which are self-contained, metadata-enriched semantic units derived from the interaction history~\cite{xu2025amem}. The Context Manager organizes these notes into a fixed Core Memory containing mission information, a Local Note Queue, and a peer-specific Neighbor Note Queue. The queues are token-bounded to prevent uncontrolled context growth while preserving information relevant to subsequent reasoning.

For information dissemination, we draw on a fundamental property of effective gossip: information should be forwarded only when it is novel to the recipient. Unlike conventional gossip mechanisms that are agnostic to message content, our approach evaluates forwarding decisions directly from the semantic content of the information being exchanged. For each one-hop neighbor, newly generated or received notes are compared with the peer-specific history using semantic similarity and recency. A strong match with recent information indicates that the peer likely already possesses equivalent knowledge, whereas semantic dissimilarity indicates greater novelty. Fixed thresholds map this novelty estimate to suppression, compact forwarding, or full forwarding. These decisions are performed deterministically rather than through additional SLM invocations, providing low-overhead and transparent information dissemination.
\newpage
The main contributions of this work are: 
\begin{itemize} 
\item \textbf{Distributed agent lifecycle:} mission-agnostic lifecycle for local UAV control by independent SLM-based agents, without a centralized coordinator;
\item \textbf{Context organization:} A bounded memory organization that structures mission, local, and peer-specific knowledge as semantic notes for long-running SLM control; 
\item \textbf{Interest-Aware Gossip Engine:} A deterministic recipient-specific dissemination policy that combines semantic novelty and recency to suppress, compact, or fully forward information. 
\end{itemize}

We evaluate the architecture using ten UAVs in a search-and-rescue mission involving 5, 10, or 20 moving survivors, with the objective of correctly identifying the total number of individuals. The proposed policy completes all experimental runs, whereas the evaluated SLM-based baselines exhibit mission failures. Compared with the most successful SLM baseline, the proposed approach uses approximately half the inference tokens and reduces transmitted data from approximately 6--7~MB to below 0.5~MB per mission, while achieving a lower survivor-count error.

The remainder of this paper is organized as follows. Sec.~\ref{sec:rw} reviews related work; Sec.~\ref{sec:architecture} presents architecture; Sec.~\ref{sec:memory} and Sec.~\ref{sec:interest_score} describe context management and interest-aware gossip, respectively; Sec.~\ref{sec:performance} reports the evaluation; and Sec.~\ref{sec:conclusion} concludes the paper.

%% file: related.tex
\section{Related Works}
\label{sec:rw}
Early work on embodied control focused on grounding language-model outputs into executable actions. For example, language-model skill scores have been combined with learned affordances to select feasible robot actions, while robot-specific language agents have been used for collaborative planning with centralized
motion planning~\cite{Mandi2023RoCo:}. 
More recent UAV-oriented work has explored different ways of placing and using language models, including centralized language interpretation with distributed control, edge-assisted architectures with lightweight per-UAV reasoning, and separate language-model decision processes for individual UAVs
~\cite{Ahmed2026LLM-Guided,Wang2025RALLY:}. In these systems, language models are mainly used for mission-level reasoning, task planning, or navigation decisions, while lower-level control is handled by conventional controllers.

When multiple agents cooperate, a key challenge is deciding whether, what, and with whom to communicate. Selective-communication methods address this by identifying decision-relevant neighbors, learning compact messages under communication budgets, or selecting human-readable state features for multi-robot coordination~\cite{2021Learning,Su2026Towards}. Despite different message representations, they share the goal of avoiding communication that does not aid coordination. Semantic and significance-aware communication follows the same principle by considering the value of information for a specific task, time, or receiver~\cite{Uysal2021Semantic}.

Gossip and peer-to-peer protocols extend information sharing beyond direct agent pairs. Gossip research has mainly studied freshness, redundancy, topology,
and repeated local forwarding, while semantic and multi-objective dissemination has also been identified as an important research direction~\cite{10893697}.
Semantic peer-to-peer approaches have also combined embeddings and interest-based organization to route events toward relevant peers~\cite{Zaarour2022OpenPubSub:}. In UAV networks, adaptive gossip protocols have instead used packet-reception history to change forwarding probabilities
and neighbor selection~\cite{Ren2024Intelligent}. These works show how multi-hop dissemination can reduce indiscriminate flooding through semantic routing, selective forwarding, and knowledge of previous transmissions.

These research lines provide main building blocks, but are addressed in isolation. In contrast, our work integrates fully distributed UAV control, semantic memory, and information dissemination into a unified multi-agent architecture in which each UAV hosts an independent SLM. 

%% file: arch.tex
\section{System Model and Architecture}
\label{sec:architecture}

We consider a mission-agnostic architecture in which UAV behavior is specified through the agent context, while vehicle interaction and information dissemination remain independent of the particular mission. This section introduces the system model and per-UAV architecture, while the memory-management and interest-aware gossip mechanisms are detailed in the Sec~\ref{sec:memory} and Sec~\ref{sec:memory}, respectively.
\newpage
\subsection{System Model}
\label{subsec:system_model}

Formally, we consider a swarm of $N$ UAVs, $\mathcal{U}=\{U_1,\ldots,U_N\}$, that cooperatively execute a common mission. Each UAV $U_i$ hosts an independent SLM-based agent $A_i$, together with its local execution context and memory. The internal state of each agent evolves according to its local observations, executed actions, and information received from other UAVs. 
We target distributed missions in which each UAV observes only part of the environment and maintains its own view of the mission. UAVs can exchange mission-relevant information with nearby peers, and such information can propagate over multiple hops when direct communication is not available. No UAV is assumed to maintain a complete or globally synchronized view of the swarm.

Time is denoted by $t\in\mathbb{R}_{\geq 0}$. Since UAV mobility can change communication opportunities during mission execution, swarm connectivity is represented by the time-varying graph
\(
    \mathcal{G}(t)=\bigl(\mathcal{U},\mathcal{E}(t)\bigr),
\)
where $(U_i,U_j)\in\mathcal{E}(t)$ indicates that UAV $U_i$ can directly transmit information to UAV $U_j$ at time $t$. The one-hop neighbor set of $U_i$ is therefore:
\begin{equation}
    \mathcal{N}_i(t)=
    \left\{
        U_j\in\mathcal{U}\setminus\{U_i\}
        \mid (U_i,U_j)\in\mathcal{E}(t)
    \right\}.
\end{equation}
No specific topology is assumed. Information may be exchanged directly with current neighbors and subsequently relayed across multiple hops. 

\subsection{Per-UAV Architecture and System Lifecycle}
\label{subsec:per_uav_architecture}

\begin{figure}
    \centering
    \includegraphics[width=0.75\linewidth]{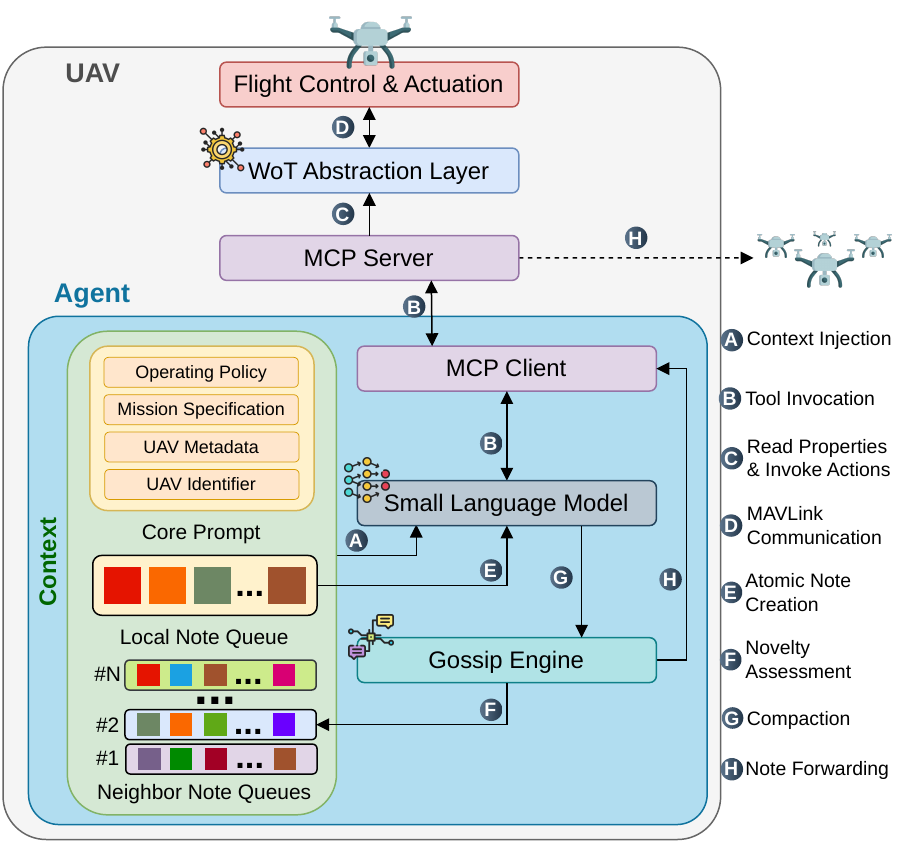}
    \caption{Per-UAV architecture and agent lifecycle.}
    \label{fig:architecture}
    \vspace{-1em}
\end{figure}
Fig.~\ref{fig:architecture} presents the architecture instantiated onboard each UAV. Its operation follows an event-driven reason--act--observe lifecycle centered on Model Context Protocol (MCP) interactions that expose vehicle capabilities to the agent. At mission initialization, the mission specification, operating policy, UAV metadata, and descriptions of the available capabilities form the initial agent context. At each reasoning step, the SLM interprets the current context and selects a mission-level action together with its input parameters. The action is issued as a structured direct MCP tool invocation (\textit{and not code generation}) and executed by the UAV. Its outcome---whether a successful result, an observation, or an error---is returned through an MCP callback, incorporated into the context, and triggers the next SLM invocation. This cycle continues until the mission is completed, determined to be infeasible under the current conditions (and then a Return to Launch command is issued), or explicitly terminated. Messages received asynchronously from other UAVs become available at the next reasoning step, allowing subsequent actions to account for peer knowledge without a separate synchronization phase. 

The interaction path follows the Web of Things (WoT) abstraction introduced in~\cite{iannoli2026say}, selected for its natural-language expressivity and its ability to abstract heterogeneous hardware devices behind a common set of invocation functions. The MCP client is the agent-side interface that conveys structured tool requests and their results. The local MCP server acts as a gateway between the agent and the UAV, which exposes UAV capabilities as MCP tools and maps each request to a WoT property read or action invocation. The WoT Abstraction Layer describes these capabilities through platform-independent WoT interfaces and translates their use into operations supported by the underlying vehicle interface, represented by MAVLink in Fig.~\ref{fig:architecture}. This separation allows the SLM to reason over semantic capabilities such as navigation or sensing without producing code or manipulating platform-specific flight commands.

The labeled flows in Fig.~\ref{fig:architecture} show how this lifecycle is realized. Before each reasoning step, a bounded context containing core mission information and selected local and peer notes is provided to the SLM (\textbf{A}). The resulting tool invocation passes through the MCP client to the MCP server (\textbf{B}). The server accesses the requested property or invokes the selected action through the WoT Abstraction Layer (\textbf{C}), which communicates with the flight-control stack through MAVLink (\textbf{D}). Execution results return along the same path and initiate the next reasoning step. 
Each completed SLM invocation and its associated execution feedback produce an atomic note (\textbf{E}). Locally generated notes are stored in the Local Note Queue, whereas received notes retain their source association in the corresponding Neighbor Note Queue. Together with the fixed core information, these bounded queues determine what is available during subsequent context injections. Their representation and retention rules are presented in Sec.~\ref{sec:memory}. 
Atomic notes are processed by the Interest-Aware Gossip Engine. For every candidate note, the engine algorithmically evaluates its novelty for each current neighbor using the knowledge associated with that peer (\textbf{F}). The result determines whether the note is suppressed, forwarded in full, or compacted. Only when compaction is selected is the SLM invoked to generate a shorter semantic representation (\textbf{G}). The selected representation is then transmitted to the corresponding neighbor (\textbf{H}). A received note is added to the recipient's peer-specific context and may subsequently be considered for further dissemination, enabling mission knowledge to propagate across swarm topology utilizing the gossip engine described in Sec.~\ref{sec:interest_score}.

\section{Context and Memory Management}
\label{sec:memory}

The continuous agent lifetime described in Section~\ref{subsec:per_uav_architecture} produces an evolving stream of observations, actions, execution feedback, and information received from other UAVs. Providing the complete history to the SLM at every invocation would progressively increase the context size and mix information with different roles in the reasoning process. The Context Manager addresses this problem by
organizing persistent information into separate memory areas and by constructing a bounded context for each SLM invocation.

The proposed organization follows the general idea of separating active and persistent information in language-model agents~\cite{packer2023memgpt}, while using structured atomic notes to represent individual pieces of semantic knowledge of A-Mem~\cite{xu2025amem}. In our architecture, however, the memory is organized specifically around the operation of a distributed UAV agent: persistent mission information is separated from knowledge acquired locally and from knowledge associated with cooperation with other UAVs. This separation supports both local reasoning and the communication decisions performed by the Gossip Engine.

\subsection{Atomic Note Construction}
\label{subsubsec:atomic_notes}

Runtime information is represented as structured \emph{atomic notes}, each containing one self-contained piece of mission-relevant knowledge. Atomic notes enter the Context Manager through two paths. First, each response returned by a tool invoked following an SLM decision is converted into a new atomic note. Second, messages received from neighboring UAVs already contain atomic notes; therefore, the received notes are stored directly rather than reconstructed, preserving their identifiers and semantic contents. We denote a generic atomic note by $m_x$, where $x$ identifies the note:
\begin{equation}
    m_x =
    \langle
        h_x,u_x,t_x,c_x,r_x,\lambda_x,\mathcal{K}_x,\Gamma_x
    \rangle,
\label{eq:atomic_note}
\end{equation}
where $h_x$ is a unique note identifier, $u_x$ identifies the UAV that originally generated the information, and $t_x$ is the time at which the represented information was generated or observed. The field $c_x$ contains the factual semantic content, whereas $r_x$ provides a concise textual summary. The set $\mathcal{K}_x$ contains keywords describing the main concepts represented by the note, and $\Gamma_x$ collects optional application-specific metadata, such as the spatial position, source type, or importance of the observation. The tag $\lambda_x$ identifies the semantic category of the note and is assigned automatically according to the type and outcome of the interaction that produced it, e.g., a successful execution of a movement command is assigned the \texttt{ROUTE\_CONTROL} tag.

\subsection{Memory Organization}
\label{subsubsec:memory_organization}

For UAV $U_i$, the Context Manager maintains three logical memory areas: the \emph{Core Memory}, the \emph{Local Note Queue}, and the \emph{Neighbor Note Queue} (one queue per peer). 
The \emph{Core Memory} contains the static information that defines the agent and its mission. As shown in Fig.~\ref{fig:architecture}, it is implemented through the Core Prompt and includes the operating policy, mission specification, UAV metadata, and UAV identifier. It is initialized at mission start, remains unchanged during execution, and is included in every SLM invocation. 
The \emph{Local Note Queue} stores the atomic notes produced by $U_i$ for its own reasoning. These notes include locally generated tool responses. Let $\operatorname{tok}(m)$ denote the number of model-input tokens required to represent note $m$, and let $B_{\mathrm{L}}$ denote the token capacity of the Local Note Queue. The queue is represented as:
\begin{equation}
    \mathcal{M}_i(t)
    =
    \left[
        m_i^{(1)},\ldots,m_i^{(L_i(t))}
    \right],
    \qquad
    \sum_{\ell=1}^{L_i(t)}
    \operatorname{tok}\!\left(m_i^{(\ell)}\right)
    \leq B_{\mathrm{L}},
    \label{eq:local_memory}
\end{equation}
where $L_i(t)$ is the number of notes stored at time $t$.

The \emph{Neighbor Note Queue} is organized separately for each peer. For every UAV $U_j$ with which $U_i$ exchanges information, the Context Manager maintains the notes received: 
\begin{equation}
    \mathcal{H}_{i,j}(t)
    =
    \left[
        m_{i,j}^{(1)},\ldots,m_{i,j}^{(L_{i,j}(t))}
    \right],
    \qquad
    \sum_{\ell=1}^{L_{i,j}(t)}
    \operatorname{tok}\!\left(m_{i,j}^{(\ell)}\right)
    \leq B_{\mathrm{N}},
    \label{eq:peer_history}
\end{equation}
where $L_{i,j}(t)$ is the number of notes currently stored for peer $U_j$, and $B_{\mathrm{N}}$ is the token capacity independently assigned to each Neighbor Note Queue. The queue stores the atomic notes received by $U_i$ from $U_j$. The same value of $B_{\mathrm{N}}$ is used for all neighbors. 

Both note queues are ordered from oldest to newest and enforce their token budget independently. Whenever inserting a new note would cause a queue to exceed this budget, its oldest notes are removed until the constraint is restored.

The semantic tag assigned during note construction enables category-specific access to both queues. Given a tag $\lambda$, we define:
\begin{equation}
    \mathcal{M}_i^\lambda(t)
    =
    \left\{
        m_x \in \mathcal{M}_i(t)
        \mid \lambda_x = \lambda
    \right\}
\label{eq:local_memory_tag}
\end{equation}
as the subset of the Local Note Queue containing notes associated with $\lambda$. Similarly:
\begin{equation}
    \mathcal{H}_{i,j}^\lambda(t)
    =
    \left\{
        m_x \in \mathcal{H}_{i,j}(t)
        \mid \lambda_x = \lambda
    \right\}
\label{eq:peer_history_tag}
\end{equation}
denotes the corresponding subset of the Neighbor Note Queue maintained by $U_i$ for $U_j$.

At each SLM invocation, the Context Manager constructs the input context from the static Core Memory, the current Local Note Queue, and the current contents of all Neighbor Note Queues. The token limits imposed on the note queues prevent the runtime portion of this context from growing without bound.

%% file: gossip.tex
\section{Interest-Aware Gossip Engine}
\label{sec:interest_score}

The Interest-Aware Gossip Engine follows a simple principle: \textit{information should be forwarded to a peer only when it is sufficiently novel with respect to what that peer is expected to already know}. Whenever a new atomic note is generated or received, the engine evaluates it independently for each one-hop neighbor by comparing its semantic content with notes carrying the same tag in the Neighbor Note Queue. Similar and recent notes provide stronger evidence that the candidate information is already known to the neighbor, whereas dissimilar information indicates novelty. Only notes belonging to mission-relevant semantic categories, predefined before execution according to the mission objectives, are evaluated for forwarding. Based on the resulting novelty estimate, the engine either suppresses the transmission, sends a compacted representation, or forwards the complete atomic note.

In detail, the forwarding process is triggered
whenever UAV~$U_i$ generates or receives an atomic note that is admitted to its
local memory. At evaluation time $t$, a candidate note is considered
independently for each current one-hop neighbor
$U_j\in\mathcal{N}_i(t)$ that is not known to have already received the same
note. 
Each transmitted note is accompanied by forwarding metadata, including the
previous sender, ordered path, hop count, gossip start time, transmission time,
and time-to-live (TTL). This metadata is maintained separately from the atomic
note and does not modify its semantic content. 
A received duplicate note is identified through its note identifier and
discarded.

For each considered neighbor $U_j$, the engine estimates the semantic novelty
of the candidate note with respect to the peer-specific history maintained by
UAV~$U_i$. Since $U_i$ cannot directly observe the internal memory of $U_j$, it
uses $\mathcal{H}_{i,j}(t)$ as its local estimate of the information already
available to that peer. For a candidate note $m_n$ with semantic tag
$\lambda_n$, the comparison is restricted to the corresponding tag-specific
peer history $\mathcal{H}_{i,j}^{\lambda_n}(t)$.
Each atomic note $m_x$ is mapped to a dense semantic representation through the
text encoder $f_{\mathrm{enc}}$:
\begin{equation}
    \mathbf{e}_x =
    f_{\mathrm{enc}}
    \left(
        c_x
        \mathbin{\Vert}
        r_x
        \mathbin{\Vert}
        \mathcal{K}_x
        \mathbin{\Vert}
        \lambda_x
    \right),
    \label{eq:note-embedding}
\end{equation}
where $\Vert$ denotes textual concatenation. 
Thus, $\mathbf{e}_n$ denotes the embedding of the candidate note, while
$\mathbf{e}_\ell$ denotes the embedding of a historical note
$m_\ell\in\mathcal{H}_{i,j}^{\lambda_n}(t)$.
All remaining note fields and
forwarding metadata are excluded from the embedding. Embeddings are computed
and cached locally and are not transmitted with the note.

Notes belonging to the same semantic category may contain recurrent vocabulary
and structural patterns. As a result, their raw embeddings can exhibit a
systematic similarity even when the represented information is different. To
reduce this tag-dependent component, UAV~$U_i$ maintains a mean embedding for
each semantic tag. Using the tag-specific local memory
$\mathcal{M}_i^\lambda(t)$, the mean embedding for tag $\lambda$ is:
\begin{equation}
    \overline{\mathbf{e}}_{i}^{\lambda}(t)
    =
    \frac{1}{\left|\mathcal{M}_{i}^{\lambda}(t)\right|}
    \sum_{m_q\in\mathcal{M}_{i}^{\lambda}(t)}
    \mathbf{e}_q,
    \label{eq:tag-mean-embedding}
\end{equation}
for $\left|\mathcal{M}_{i}^{\lambda}(t)\right|>0$. These mean embeddings are
cached locally and updated as the corresponding tag-specific memory changes.

For the candidate note $m_n$ and a historical note
$m_\ell\in\mathcal{H}_{i,j}^{\lambda_n}(t)$, the corresponding residual
embeddings are obtained by subtracting the same tag-level mean:
\begin{equation}
    \widehat{\mathbf{e}}_n
    =
    \mathbf{e}_n
    -
    \overline{\mathbf{e}}_{i}^{\lambda_n}(t),
    \qquad
    \widehat{\mathbf{e}}_\ell
    =
    \mathbf{e}_\ell
    -
    \overline{\mathbf{e}}_{i}^{\lambda_n}(t).
    \label{eq:embedding-residuals}
\end{equation}
The residual representations capture the deviation of each note from the typical semantic content associated with tag $\lambda_n$. When fewer than five notes with tag $\lambda_n$ are available in the local memory, the tag-level correction is not applied and the original embeddings
are used, i.e., $\widehat{\mathbf{e}}_n=\mathbf{e}_n$ and
$\widehat{\mathbf{e}}_\ell=\mathbf{e}_\ell$. Then, we utilize the nonnegative cosine similarity (to keep the scores bounded in the $[0,1]$ interval) to calculate the similarity between them: 
\begin{equation}
\operatorname{cos}_{+}(\widehat{\mathbf{e}}_n,\widehat{\mathbf{e}}_{\ell})
=
\max
\left\{
0,
\frac{\widehat{\mathbf{e}}_n^{\mathsf T}\widehat{\mathbf{e}}_{\ell}}
{\lVert\widehat{\mathbf{e}}_n\rVert_2\lVert\widehat{\mathbf{e}}_{\ell}\rVert_2}
\right\}.
\label{eq:nonnegative-cosine}
\end{equation}
 At evaluation time $t$, the recency weight of historical note
$m_\ell$ is:
\begin{equation}
w_\ell(t)
=
\max
\left\{
0,
1-\frac{t-t_\ell}{W}
\right\},
\qquad t \geq t_\ell,
\label{eq:recency-weight}
\end{equation}
where $W$ is the recency window and determines how quickly a historical note loses influence. The novelty contribution of an entry is calculated as:
\begin{equation}
d_\ell=1-\operatorname{cos}_{+}(\widehat{\mathbf{e}}_n,\widehat{\mathbf{e}}_{\ell})w_\ell(t).
\label{eq:knownness-novelty}
\end{equation}
A recent and semantically similar entry provides strong evidence that the neighbor already knows the candidate information, resulting in a low $d_\ell$. Older or semantically dissimilar entries produce the opposite result.

Let
$\mathcal{V}_{i,j}(m_n)=
\{d_\ell\mid m_\ell\in\mathcal{H}^{\lambda_n}_{i,j}\}$
denote the set of novelty contributions computed between candidate note $m_n$ and the entries of neighbor~$j$'s filtered history. The neighbor-specific novelty score is:
\begin{equation}
N_{i,j}(m_n)
=
\min_{d\in\mathcal{V}_{i,j}(m_n)} d.
\label{eq:peer-novelty-score}
\end{equation}
If $\mathcal{V}_{i,j}(m_n)$ is empty, $N_{i,j}(m_n)$ is set to $1$. We utilize the minimum value since the forwarding decision is based on the most similar note. Consequently, a value near zero indicates that the neighbor is expected to know the information, whereas a value near one indicates that the information is likely to be novel. The $N_{i,j}(m_n)$ novelty value is used directly as the forwarding score and a decision is made to forward, compact or disregard the note based on previously set tag-specific thresholds ($\tau_{\mathrm{send}}^{\lambda_n}$ and $\tau_{\mathrm{full}}^{\lambda_n}$, which are determined offline by the system administrator for each tag $\lambda_n$).  The forwarding action from UAV~$i$ to neighbor~$j$ is:
\begin{equation}
a_{i\rightarrow j}(m_n)
=
\begin{cases}
\textsc{Discard},
&
N_{i,j}(m_n)
<
\tau_{\mathrm{send}}^{\lambda_n},
\\[1mm]
\textsc{Compact},
&
\tau_{\mathrm{send}}^{\lambda_n}
\leq
N_{i,j}(m_n)
<
\tau_{\mathrm{full}}^{\lambda_n},
\\[1mm]
\textsc{Full},
&
N_{i,j}(m_n)
\geq
\tau_{\mathrm{full}}^{\lambda_n}.
\end{cases}
\label{eq:gossip-forwarding-decision}
\end{equation}

A \textsc{Discard} decision suppresses transmission only to the evaluated neighbor and does not remove the note from local memory. A \textsc{Compact} decision transmits a reduced representation of the note using the local SLM for summarization. A \textsc{Full} decision transmits the complete atomic note.

%% file: eval.tex
\section{Performance Analysis}
\label{sec:performance}
We evaluate the proposed approach in a simulated multi-UAV disaster-recovery scenario in which the swarm must locate and count moving survivors. We use this scenario as an illustrative case study to evaluate the proposed architecture rather than as an optimized solution to the search-and-rescue task itself. The task requires autonomous reasoning and inter-UAV coordination, as individual UAVs must exchange observations and determine whether a detected person has already been identified elsewhere in the swarm. Mission completion measures the ability of the overall agentic control architecture to sustain coordinated execution throughout the mission workflow, whereas survivor-count accuracy quantifies the effectiveness of collaborative information sharing and identity reconciliation. Together, these metrics provide complementary measures of the effectiveness of the proposed architecture. The experimental setup is described in Sec.~\ref{sub:experimental-setup}, while the results are reported in Sec.~\ref{sub:results}.

\subsection{Experimental Setup}
 \label{sub:experimental-setup}
The experimental platform comprises 10 UAVs simulated using ArduPilot software-in-the-loop (SITL)\footnote{https://ardupilot.org/dev/docs/sitl-simulator-software-in-the-loop.html}. Each UAV is represented by an independent ArduPilot multicopter instance, with commands and telemetry exchanged through MAVLink and exposed to higher software layers through an HTTP bridge and a dedicated WoT interface (as in \cite{iannoli2026say}). Each UAV also interacts with an independent MCP server, emulating the local MCP deployment used by the proposed architecture. UAVs operate under clear and stable weather conditions, at a fixed altitude of 15~m and with a cruise speed randomly selected between 13 and 16~m/s. No UAV has access to the Internet or to a base station; all computation is performed locally.

Inter-UAV communication is analytically emulated using empirical parameters derived from IEEE 802.11n UAV measurements~\cite{10.1145/2505469.2505472}. The simulator continuously updates directed links according to UAV positions and combines a distance-dependent empirical throughput model with an analytical radio model to determine link connectivity and capacity. 

The mission area spans $200\times400$~m and is divided into ten adjacent $20\times400$~m strips, one per UAV. Each UAV first moves to the entrance of its assigned strip, announces its readiness, and waits until readiness information from the entire swarm has been received. It then traverses the strip while sampling a $20\times20$~m sensing footprint every 0.5~s. A survivor is detected whenever its simulated position lies within this footprint. Although the synchronization procedure is specified in the agent prompt and supported by MCP tools, the SLM remains responsible for invoking the required actions and may deviate from the prescribed sequence. A mission is considered completed only if all UAVs complete scanning, execute the recovery procedure, land, and enter the disarmed state. 

We evaluate scenarios containing 5, 10, and 20 survivors. At the beginning of each run, survivors are assigned random initial positions and walking/running speeds between 0.9 and 3.4~m/s and subsequently follow a random-walk mobility model. Consequently, individuals may move across strip boundaries and be detected by multiple UAVs, requiring the swarm to reconcile repeated observations.

We assume that person detection and segmentation are performed by a standard embedded computer-vision model, such as YOLOv10~\cite{wang2024yolov10}, and consider this processing stage outside the scope of this work. Person images are drawn from the UAV-Human dataset~\cite{uav-human}, which provides cropped aerial observations acquired from UAV viewpoints. To reproduce variability in natural-language descriptions, each image is processed multiple times before the experiments by the same model used by the UAV agents, using an identical prompt and a temperature of 0.8. During a mission, one description is randomly selected for each detected survivor and provided to the corresponding UAV agent. The SLM then determines, from the current description and previously observed individuals, whether the detection corresponds to a known survivor or to a new one. Each UAV maintains a local count of the survivors it identifies as new based on its locally available knowledge, including information received from other UAVs. A detection associated with an already known survivor does not increase this count. At mission completion, the local counts are summed to obtain the mission-level survivor estimate.

Each UAV agent uses Gemma 4 26B A4B IT~\cite{team2026gemma}. We selected this model because it supports deployment on edge platforms such as NVIDIA Jetson~\cite{gemma4_jetson} and, through its mixture-of-experts architecture, activates only approximately 3.8B parameters per inference step, reducing the computational overhead and latency associated with local reasoning.
\newpage
We compare the proposed approach against three policies while keeping the mission, communication environment, and agent configuration unchanged:
\begin{itemize}
\item \textit{Deterministic}: a deterministic non-cooperative controller, rather than an SLM, executes the mission workflow; i.e., launch, synchronization, and strip scanning. At the end of the mission, the survivor counts independently obtained by each UAV are summed to produce the final estimate. 

\item \textit{Full Flood}: each UAV is controlled by an SLM, and every operational event is forwarded in full to all one-hop neighbors and subsequently relayed through the swarm. Received messages are directly appended to the local agent context, without the context organization reported in Sec.~\ref{sec:memory}.

\item \textit{SLM Forwarding}: each UAV is controlled by an SLM, as in \textit{Full Flood}, but forwarding decisions are delegated to the model. For every candidate event and one-hop neighbor, the SLM receives the observable transcript and returns either \texttt{FORWARD} or \texttt{DISCARD}.
\end{itemize}

All SLM-based policies use the agent lifecycle described in Sec.~\ref{sec:architecture}, as it provides the common execution framework for SLM-controlled missions. Furthermore, they use the same mission prompt, MCP interfaces, and safety mechanisms. Therefore, differences among these policies primarily reflect their information-management strategies. \textit{Full Flood} represents unrestricted information dissemination, whereas \textit{SLM Forwarding} evaluates whether forwarding decisions can be delegated directly to the SLM. The deterministic baseline instead provides a reference for quantifying the operational overhead introduced by agentic reasoning.

For the proposed policy, semantic similarity is computed using the INT8-quantized \texttt{BAAI/bge-base-en-v1.5} embedding model~\cite{xiao2023cpack}, a 109-million-parameter BERT-based encoder. 
Because the gossip mechanism depends on message-history distributions, we perform ten preliminary runs to calibrate its forwarding thresholds. The resulting fixed thresholds, $(\tau_{\mathrm{send}},\tau_{\mathrm{full}})$, are set to 0.15 and 0.66, respectively, and are applied across all semantic tags. The token capacity $B_{\mathrm{L}}$ of the Local Note Queue is set to 8,192 tokens, while the capacity $B_{\mathrm{N}}$ of each Neighbor Note Queue is set to 4,096 tokens. Under this configuration, the maximum agent context is 45,056 tokens, well below the 256K-token context window of Gemma 4 26B A4B. The temporal window $W$ is set to 900~s.
Finally, each experimental configuration is repeated 20 times using fixed random seeds, and 90\% confidence intervals are reported.

\subsection{Results}
\label{sub:results}
\begin{table}[t]
\caption{mission completion by policy and scenario size (20 replications).}
\centering
\label{tab:success-rate}
\begin{tabular}{lccc}
\toprule
Policy & 5 people & 10 people & 20 people \\
\midrule
Full flood       & 85\% & 85\% & 70\% \\
SLM forwarding   & 0\%  & 0\%  & 0\%  \\
Deterministic    & 100\% & 100\% & 100\% \\
\midrule
\textbf{Proposal}    & 100\% & 100\% & 100\% \\
\bottomrule
\end{tabular}
\vspace{-1.5em}
\end{table}
Tab.~\ref{tab:success-rate} shows that the deterministic baseline and the proposed policy achieved operational completion in all mission runs. In contrast, \textit{Full Flood} achieved completion rates between 70\% and 85\%, whereas \textit{SLM forwarding} failed to complete any run.  Failed \textit{Full Flood} runs generally reached the final mission phase but did not complete one or more required recovery actions before termination (e.g., landing or disarming). These failures are consistent with context degradation caused by unrestricted message replication, which progressively enlarges the SLM context with redundant and heterogeneous observations and may reduce its ability to prioritize terminal safety actions. The \textit{SLM forwarding} policy instead failed during the initial synchronization phase in all replications, with a median of only 4 out of 10 UAVs reaching the ready state. Because every candidate event was independently evaluated for each one-hop neighbor, forwarding decisions generated additional reasoning steps and events, producing a cascading overhead that diverted execution from flight-control actions. Each run generated an average of 1,078 forwarding decisions before termination. 
 The lack of context organization further increased the amount of heterogeneous information exposed to the SLM. The remaining results are reported only for successful runs; therefore, \textit{SLM forwarding} is excluded, while results for \textit{Full Flood} should be interpreted as conditional on successful mission completion.

\begin{figure}
    \centering
    \includegraphics[width=0.6\linewidth]{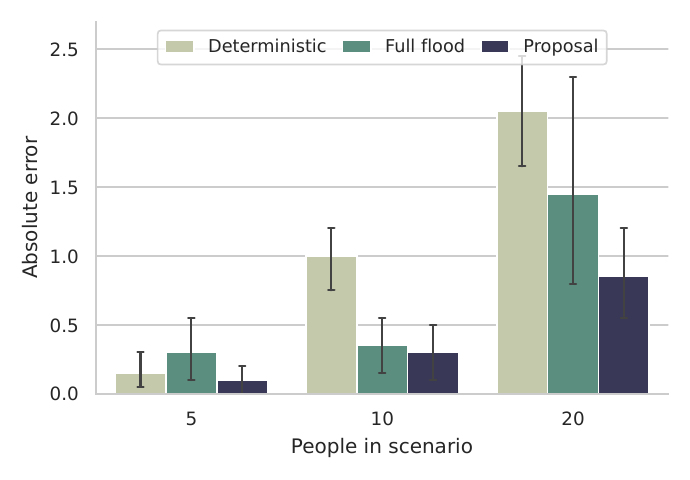}
    \caption{Average absolute count error per mission.}
    \label{fig:count-error}
    \vspace{-1.1em}
\end{figure}

Fig.~\ref{fig:count-error} reports the mean absolute survivor-count error per mission across the evaluated scenarios. As expected, the \textit{Deterministic} baseline systematically overcounts survivors, with the error increasing approximately with the number of people in the scenario. Since this policy performs no semantic reasoning or identity reconciliation, a survivor moving across search strips may be detected by multiple UAVs and consequently counted more than once. \textit{Full Flood} exhibits slightly higher errors than the proposed policy in all evaluated scenarios. Its wider confidence intervals, particularly with 20 survivors, are partly due to the smaller number of successful replications resulting from the failures reported in Tab.~\ref{tab:success-rate}. Errors in the proposed policy mainly arise from identity-association mistakes, either merging two visually similar individuals into a single survivor or treating different observations of the same survivor as distinct individuals. These errors are less frequent than with \textit{Full Flood}, as the interest-aware gossip engine limits the propagation of redundant descriptions to UAVs that already possess information about the corresponding survivor.

\begin{figure}[t]
    \centering
    \includegraphics[width=0.75\linewidth]{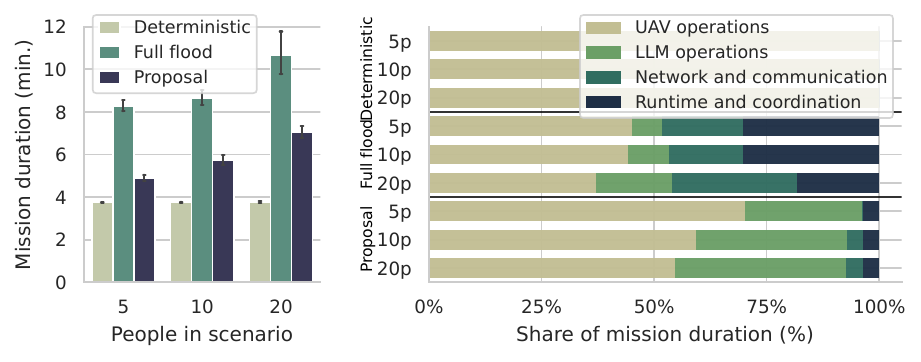}
    \caption{Average mission duration (left panel) and mission share per operation type (right panel).}
    \label{fig:duration-and-battery}
    \vspace{-1.3em}
\end{figure}

The left panel of Fig.~\ref{fig:duration-and-battery} reports the average mission duration, while the right panel reports the execution-time share of each operation category.  The \textit{Deterministic} baseline remains approximately constant across scenario sizes because it does not incur SLM reasoning or the associated message-management overhead. In contrast, both SLM-based policies require progressively more time as the number of survivors increases, since additional detections trigger further identity reasoning, communication, and message processing. \textit{Full Flood} exhibits the longest missions because unrestricted dissemination causes more notes to be forwarded, processed, and incorporated into the agents' contexts.  The right panel shows that this cost is not only due to network transmission time; a substantial fraction appears under runtime and coordination, which includes context assembly, forwarding decisions, MCP scheduling, duplicate/path/TTL checks.  Thus, the gain of our approach comes not only from lowering network traffic, but also from avoiding the execution-time cost of managing unnecessary information.

\begin{figure}
    \centering
    \includegraphics[width=0.75\linewidth]{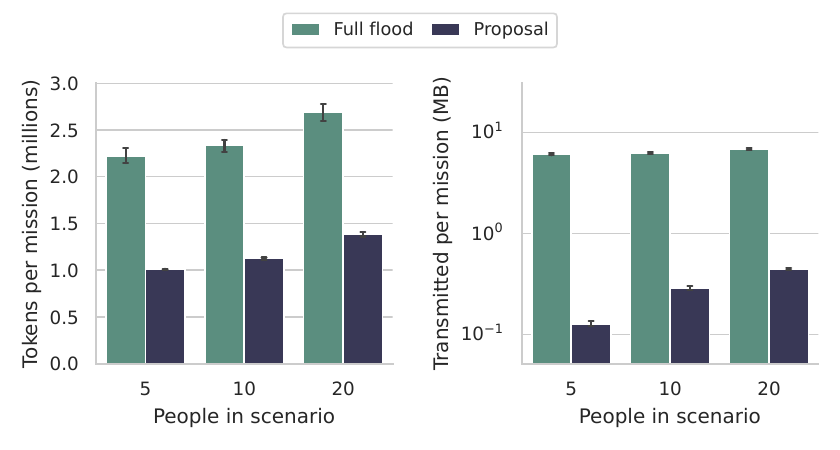}
    \caption{Average token consumption per mission (left panel) and average data transmitted per mission (right panel, y-axis on a logarithmic scale). }
    \label{fig:token-network}
    \vspace{-1em}
\end{figure}

Fig.~\ref{fig:token-network} quantifies the reasoning (left panel) and communication overhead (right panel). Token consumption increases with the number of survivors for both SLM-based policies; however, the proposed policy consistently uses approximately half as many tokens as \textit{Full Flood}. The difference is even more significant in terms of network traffic, where \textit{Full Flood} transmits approximately 6--7~MB per mission and the proposed policy remains below 0.5~MB across all evaluated scenarios.

\newpage
Overall, the results show that fully agentic UAV missions are feasible, but their effectiveness depends strongly on how reasoning and information are managed. The deterministic baseline demonstrates the value of semantic reasoning for survivor identity reconciliation. Conversely, \textit{Full Flood} shows that indiscriminate message dissemination increases reasoning, communication, and context-management overhead and can ultimately compromise mission completion. Delegating forwarding itself to the SLM is even less practical, as the resulting cascade of reasoning operations prevents timely execution of the mission workflow. Taken together, these results indicate that effective agentic swarm coordination benefits from structured context memory and selective information dissemination.

%% file: conclusions.tex
\section{Conclusion}
\label{sec:conclusion}
This paper presented a fully distributed architecture for SLM-based UAV swarms in which each UAV reasons and acts locally without relying on a centralized coordinator. The architecture combines a bounded Context Manager, which structures mission, local, and peer-specific knowledge through atomic notes, with an Interest-Aware Gossip Engine that deterministically disseminates information according to recipient-specific semantic novelty and recency. In a search-and-rescue scenario, the proposed policy consistently outperformed the evaluated SLM-based baselines in mission completion, communication overhead, inference-token consumption, and survivor-count accuracy. In particular, it completed all runs while reducing transmitted data from 6--7~MB to below 0.5~MB per mission compared with SLM baseline. The results indicate that effective agentic swarm control requires structured memory and selective dissemination, while indiscriminate flooding or delegating every forwarding decision to the SLM introduces substantial operational overhead. Future work will evaluate larger and more heterogeneous swarms, other embedded models and additional mission types.